\documentclass[fleqn,usenatbib]{mnras}

\usepackage{newtxtext,newtxmath}

\usepackage[T1]{fontenc}

\DeclareRobustCommand{\VAN}[3]{#2}
\let\VANthebibliography\thebibliography
\def\thebibliography{\DeclareRobustCommand{\VAN}[3]{##3}\VANthebibliography}

\usepackage{graphicx}	
\usepackage{amsmath}	
\usepackage{subfigure}

\title[AT2022kak]{An extremely fast fading population II dwarf nova candidate: caught spectroscopically on the rise}

\author[N. Van Bemmel et al.]{Natasha Van Bemmel$^{1,2}$,
\thanks{E-mail: nvanbemmel99@gmail.com}
Jielai Zhang$^{1}$,
Jeff Cooke$^{1,2,3}$, 
Anais M\"{o}ller$^{1,2}$, 
\newauthor
Igor Andreoni$^{4,5,6}$,
Katie Auchettl$^{7,8}$,
David Buckley$^{9,10,11}$,
Jonathan Carney$^{5}$,
Dougal Dobie$^{1,2}$, 
\newauthor
James Freeburn$^{1,2}$, 
Bruce Gendre$^{12,13}$,
Vanshika Kansal$^{1,14}$,
Itumeleng Monageng$^{10}$,
Arne Rau$^{15}$, 
\newauthor
Nikita Rawat$^{10}$,
Mark Suhr$^{1,3}$,
Edward N. Taylor$^{1}$
\\
$^{1}$Centre for Astrophysics and Supercomputing, Swinburne University of Technology, Victoria 3122, Australia \\
$^{2}$ARC Centre of Excellence for Gravitational Wave Discovery (OzGrav), Victoria 3122, Australia \\
$^{3}$Australian Research Council Centre of Excellence for All-sky Astrophysics in 3 Dimensions (ASTRO 3D)\\
$^{4}$Joint Space-Science Institute, University of Maryland, College Park, MD, USA\\
$^{5}$Department of Physics and Astronomy, University of North Carolina at Chapel Hill, Chapel Hill, NC 27599-3255, USA\\
$^{6}$Astrophysics Science Division, NASA Goddard Space Flight Center, Greenbelt, MD, USA\\
$^{7}$OzGrav, School of Physics, The University of Melbourne, Victoria 3010, Australia\\
$^{8}$Department of Astronomy and Astrophysics, University of California, Santa Cruz, CA 95064, USA\\
$^{9}$Department of Astronomy, University of Cape Town, Private Bag X3, Rondebosch, Cape Town, 7701, South Africa \\
$^{10}$South African Astronomical Observatory, PO Box 9, Observatory, Cape Town, 7935, South Africa \\
$^{11}$Department of Physics, University of the Free State, PO Box 339, Bloemfontein 9300, South Africa \\
$^{12}$The University of Western Australia, OzGrav ARC Centre of Excellence, 35 Stirling Highway, 6009 Crawley, WA, Australia \\ 
$^{13}$University of the Virgin Islands, College of Science and Mathematics, 2 John Brewer's Bay, St Thomas, 00802, VI, USA\\
$^{14}$ ARC Centre of Excellence for Dark Matter Particle Physics, Victoria 3122, Australia\\
$^{15}$Max-Planck-Institut für Extraterrestrische Physik, Gießenbachstraße, 85748 Garching, Germany\\
}

\date{Accepted XXX. Received YYY; in original form ZZZ}

\pubyear{\the\year{}}

\begin{document}
\label{firstpage}
\pagerange{\pageref{firstpage}--\pageref{lastpage}}
\maketitle

\begin{abstract}
We present AT2022kak, a rapidly evolving optical transient discovered by the KiloNova and Transients Program (KNTraP). This interesting burst exhibited extremely fast evolution, with a large amplitude blue outburst of $m >$ 3.3 in a single night, and a rapid fade back to quiescence in the following two nights. We deployed a multi-wavelength follow-up campaign, monitoring the object for the next two months, but saw no recurrent burst. Three years later, while observing to get spectroscopy of the object in quiescence, there was a new outburst, enabling the collection of time-resolved spectra of the rise and fade of the outburst. The light curve properties of the first burst and spectra of the second burst are consistent with a dwarf nova. Its fast evolving behaviour makes it one of the fastest and faintest dwarf novae observed. The estimated distance of AT2022kak from the Galactic centre is $\sim$6.6 kpc, with a scale height of $\sim$2 kpc. This scale height places it in the Galactic thick disk, where only very few dwarf novae have been found, and is therefore a potential Population II dwarf novae system.

\end{abstract}




\section{Introduction}

With the increasing capabilities of transient facilities allowing faster cadence and deeper surveys, there has been a rise in the discoveries of rapidly evolving transients. The landscape of fast optical transients has broadened with new classes being uncovered, and other previously theorised transients becoming better characterised, such as fast-evolving luminous blue optical transients \citep{Drout:2014, Arcavi:2016, Tanaka:2016, Pursiainen:2018, Prentice:2018, Rest:2018, Ho:2023}, tidal disruption events \citep[][and references therein]{Andreoni:2022, Gezari:2021}, gamma-ray burst afterglows and gamma-ray burst ``orphan'' afterglows \citep{Ho:2020, Ho:2022, Perley:2025, Srinivasaragavan:2025, Freeburn:2025}.

The KiloNova and Transients Program \citep[KNTraP;][]{VanBemmel:2025} is an optical survey with an observational strategy specifically designed to optimise for the discovery of kilonovae \citep[][and references therein]{Rastinejad:2022, Levan:2023, Abbott:2017}. This search uses the Dark Energy Camera \citep[DECam;][]{Flaugher:2015} on the 4m Victor M. Blanco Telescope at the Cerro Tololo Inter-American Observatory, for its high sensitivity, wide field of view, and fast read out and filter exchange capabilities, which allows for fast multi-filter observations. KNTraP probes the sky nightly with multiple filters, and processes the data in hours for next-day analysis of transients. This survey is currently the only nightly cadence survey sensitive to m $\sim$ 25.4, and thanks to its observing strategy, aside from kilonovae, KNTraP probes the fast transients parameter space not available to other transient surveys. 

In this work, we explore the detection of AT2022kak \citep{Zhang_kak:2022}, a bright, fast evolving, blue transient discovered by KNTraP in 2022, which rose and faded over three magnitudes across three days. Given the fast evolving nature, amplitude and colour of outburst, and observations of a  recurrent burst, we rule out the possibility of this transient being an fast-evolving blue optical transient, gamma-ray burst afterglow, or kilonova. The most likely progenitor of this recurring transient with these characteristics is a dwarf nova (DN). 

DNe are outbursts occurring in cataclysmic variable (CV) systems containing a white dwarf which is accreting matter from a secondary donor star, generally a late main sequence star \citep[see][ for a review]{Warner:2003}. Thermal instabilities and changes in viscosity within the accretion disk cause large and sudden outbursts \citep[see][]{Hameury:2020}, with outburst amplitudes of $\sim$ 5 magnitudes. There is a large variety of DN outbursts, due to differing physical properties of the systems such as the mass transfer rate, orbital period, and sizes of the WD and secondary star. There are several DN subclasses, with the three main types being SU UMa, U Gem, and Z Cam that exhibit different outburst behaviours and periodicities \citep{Osaki:1996}. SU UMa are the most commonly found DN, and exhibit both shorter `normal' and longer and larger amplitude `superoutbursts'. U Gem types have periodic outbursts, which are nearly identical to the SU UMa normal outbursts. Z Cam are very rare in comparison to the other two types, and appear similar to U Gem, however they show `standstill' periods of very low, to no variation in its magnitude post-outburst. Despite the large variety of DNe, strong correlations have been observed between their physical properties, linking characteristics such as absolute peak magnitude, orbital period, distance, and secondary star mass \citep[e.g.][]{Kukarkin:1934, Bailey:1975, Coppejans:2016, Otulakowska-Hypka:2016}. 

DNe are commonly found within the disk of our Galaxy (within 0.6 kpc from the plane of the disk), with Population I systems \citep{Lee:2011, Mateu:2018}. Only a few DNe have been discovered in the Galactic thick disk or halo, as they are observationally more difficult to find due to their faintness. These distant Galactic DNe possibly have Population II secondary stars \citep{Stehle:1997}. Population II DNe are of interest as they are rarely found observationally, due to their faintness and only a few candidates have been observed \citep{Hawkins:1987, Howell:1990, Lee:2019}. Currently, only one Galactic Population II DN candidate has been observed with photometric light curves, KSP-OT-201611a \citep{Lee:2019}. This DN has a height above the Galactic plane of 1.7 kpc, placing it in either the thick disk or halo. Theoretical models of Population II DNe expect that their accretion disks should be brighter than Population I DNe systems due to larger accretion rates \citep{Stehle:1997}. However, observationally, the opposite has been found, in particular a population of distant Population II DNe in a globular cluster have been observed to have fainter luminosities than dwarf novae with a solar metallicity secondary \citep{Edmonds:2003}. 
 
In this work, we present the observational data and analysis of AT2022kak. Section \ref{sec:observations_and_data} presents the initial optical data from the KNTraP run, the infrared (IR), optical, high energy, and spectroscopic data collected in post-burst follow-up efforts, as well as the serendipitous second burst. Analysis is done on the light curve, spectroscopic data, as well as determination of the Galactic location and distance in Section \ref{sec:analysis}. The DN and its novel behaviours are presented and summarised in Section \ref{sec:conclusion}.

\section{Observations and Data}
\label{sec:observations_and_data}

The KNTraP run collected observations from 11--22 February 2022, which observed 31 fields per night with DECam's wide field of view \citep[3 deg$^{2}$, ][]{Flaugher:2015}, in the \emph{i} and \emph{g} filters, reaching a limiting magnitude of m $\sim$ 24.5 \citep[see][for details]{VanBemmel:2025}. To find fast evolving transients, the observational data were processed nightly, for next day visual inspection. AT2022kak was identified during the observation run $\sim$1 week after peak. Rapid identification and follow-up of this candidate was delayed, as the transient did not meet the program KN search criteria.

\subsection{KNTraP Photometry}

\begin{figure*}

    \centering
    \includegraphics[width=1\linewidth]{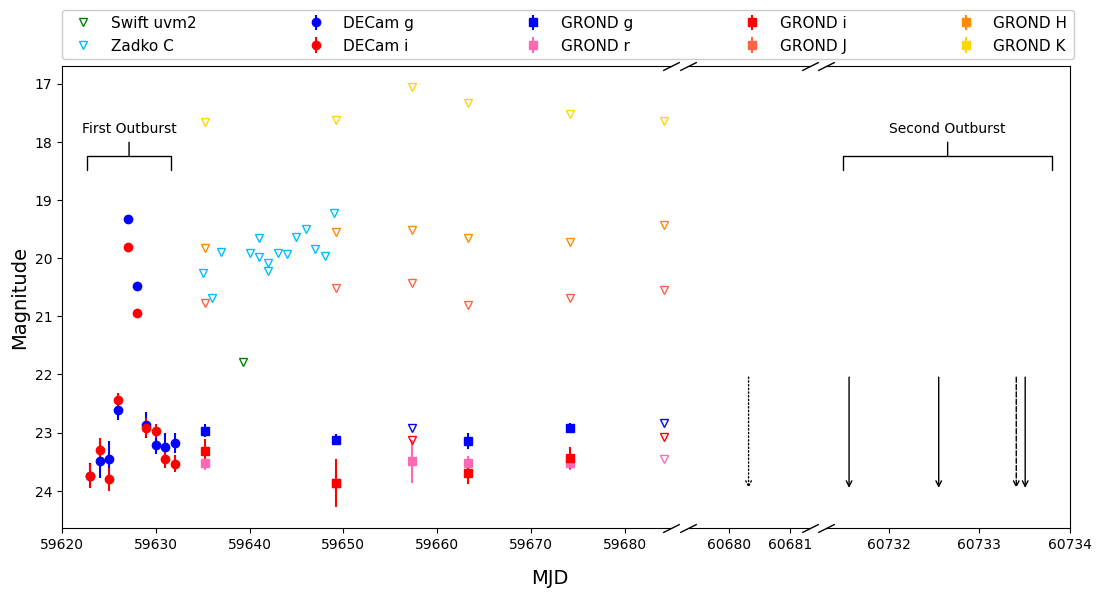}
    \caption{The optical-infrared light curve and spectroscopic follow up of AT2022kak over $\sim$3 years. KNTraP DECam \emph{i} and \emph{g} filter photometry is shown as red and blue filled circles, respectively. Follow-up photometry from GROND, Zadko, and Swift are denoted with filled squares (see legend), with upper limits shown as inverted triangles. Down-pointing arrows indicate the nights during which spectroscopy was acquired with SOAR (dotted arrow), KOALA (solid arrows), and SALT (dashed arrow). Note: light curves include quiescent source flux.}
    \label{fig:full_photometry}
\end{figure*}

The raw data collected from DECam as part of the KNTraP program were processed in batches during the night using the NOIRLab Community Pipeline \citep{Valdes:2014} and transferred to the OzSTAR supercomputer at Swinburne University of Technology and processed by {\sc Photpipe} \citep{Rest:2005, Rest:2014}. {\sc Photpipe} is an image subtraction and candidate selection pipeline, which also performs astrometric calibrations, alignment, coadditon, and photometry of the images. We first detected the transient with {\sc Photpipe} through difference imaging on 16 February 2022 at a magnitude of $m_i$ = 22.4$\pm$0.1, and reached its peak observed magnitude the following night. Within the next two nights, the burst was observed to fade from the detected peak by three magnitudes. The KNTraP light curve in included in Figure \ref{fig:full_photometry} with additional follow-up data described below, and photometry is listed in Table \ref{tab:AT2022kak_photometry}.

\begin{figure}
    \centering
    \includegraphics[width=1\linewidth]{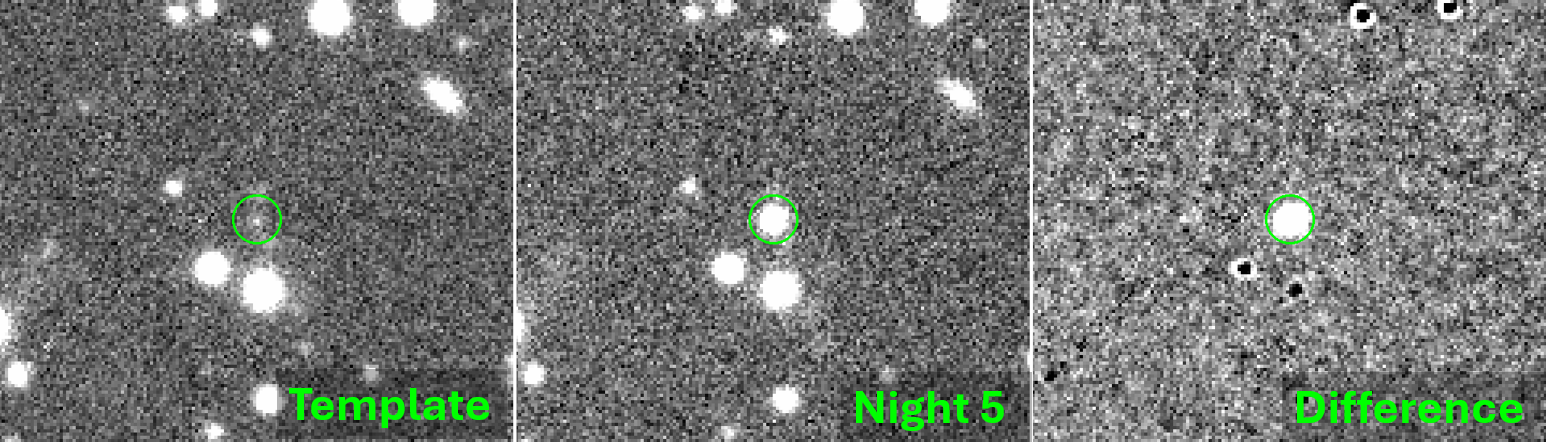}
    \caption{KNTraP images of \emph{AT2022kak} during its initial outburst. From left to right: the template image (first night's data), the science image during the peak-burst emission on night 5 (MJD 59627.27), and the difference image. The green circle indicator has a radius of 2 arcseconds.}
    \label{fig:thumbnails}
\end{figure}

\subsection{Multi-Wavelength Follow-up Observations}


\subsubsection{Optical and Infrared}

On 24 February 2022, we triggered the Gamma-ray Burst Optical/Near-infrared Detector \citep[GROND;][]{Greiner:2008} for follow-up. Observations started $\sim$8 days post-peak and detected a faint source in \emph{g, r} and \emph{i} bands with a magnitude consistent with the DECam light curve, shown in Figure \ref{fig:full_photometry}. GROND \emph{g} and \emph{i} bandpasses are similar to those in DECam. Upper limits were constrained with \emph{J, H} and \emph{K} bands. GROND continued to monitor the source weekly for $\sim$1.5 months. This monitoring shows a consistent detection of the persistent source at a magnitude of $\sim$23 (\emph{g}) and $\sim$23.5 (\emph{r,i}) and suggests that there were no recurrent bursts in this time frame. This faint persistent source can also be seen in the template image in the KNTraP DECam data in Figure \ref{fig:thumbnails}.

After 13 days post-peak, we also triggered the Zadko Observatory \citep{Coward:2017} for consistent monitoring. This telescope monitored the source each night for two months using the `clear' filter and multiple 90 second exposures per night. A nightly limiting magnitude of $m_c\sim$ 20 could be achieved through coadding the images. From these stacked images, no recurrent burst was observed. 

To attempt to characterise the quiescent source, we create a spectral energy distribution (SED) using both KNTraP photometry and archival DECam images. Photometry was also collected from the DELVE catalog in the \emph{g} and \emph{r} bands, and GROND follow-up data in \emph{g, r} and \emph{i} bands. This SED with the combined photometric data is shown in Appendix \ref{appsec:at2022kak_sed}. However, as different instruments were used for these observations, and each were taken at different times (archival DECam data was taken from years 2014 and 2019), there is up to a $\sim$1 magnitude difference between the detections in the same band that may reflect source variability.

\subsubsection{High Energy}

While the identity of the source was still unknown, we collected observational data from the Neil Gehrels Swift Observatory \citep[{\it Swift};][]{Gehrels:2004} in gamma-ray, x-ray, and ultraviolet on 1 March 2022.  

There was no detection in the {\it Swift} Ultraviolet/Optical Telescope (UVOT) data to $m_{uvm2}$ = 21.78 AB and the X-ray Telescope (XRT) to 0.3--10 keV 3.6 $\times10^{-14}$ erg/cm$^{2}$/s. As the main aim of the KNTraP search was to detect KNe, the Swift data were interpreted in the context of a KN candidate. Given the peak luminosity of a GW170817-type event at a redshift of z $\sim$ 0.03, the 3$\sigma$ X-ray luminosity upper limit corresponds to 8.1$\times10^{40}$ erg/s.

\begin{table}
    \centering
    \caption{Photometry table for AT2022kak.}
    \tabcolsep=0.11cm
    \begin{tabular}{|c|c|c|c|c|c|}
        \hline
        MJD & Phase & Filter & \emph{m} & $\delta$\emph{m} & Telescope \\
        \hline
        59623.28 & -3.99 &    i &     23.75 &     0.21 & DECam \\
        59623.28 & -3.99 &    g &     23.74 &     0.21 & DECam \\
        59624.27 & -3.00 &    i &     23.31 &     0.21 & DECam \\
        59624.28 & -3.00 &    g &     23.48 &     0.30 & DECam \\
        59625.27 & -2.00 &    i &     23.81 &     0.20 & DECam \\
        59625.28 & -2.00 &    g &     23.46 &     0.31 & DECam \\
        59626.27 & -1.00 &    i &     22.44 &     0.12 & DECam \\
        59626.28 & -1.00 &    g &     22.61 &     0.17 & DECam \\
        59627.27 &  0.00 &    i &     19.81 &    0.012 & DECam \\
        59627.27 &  0.00 &    g &     19.32 &    0.011 & DECam \\
        59628.27 &  1.00 &    i &     20.94 &    0.025 & DECam \\
        59628.27 &  1.00 &    g &     20.48 &    0.023 & DECam \\
        59629.28 &  2.01 &    i &     22.92 &     0.16 & DECam \\
        59629.28 &  2.01 &    g &     22.87 &     0.21 & DECam \\
        59630.28 &  3.01 &    i &     22.97 &     0.12 & DECam \\
        59630.28 &  3.00 &    g &     23.21 &     0.17 & DECam \\
        59631.27 &  4.00 &    i &     23.45 &     0.15 & DECam \\
        59631.27 &  4.00 &    g &     23.25 &     0.23 & DECam \\
        59632.27 &  5.00 &    i &     23.54 &     0.15 & DECam \\
        59632.27 &  5.00 &    g &     23.18 &     0.17 & DECam \\
        59635.26 &  7.99 &    g &     22.97 &     0.11 & GROND \\
        59635.26 &  7.99 &    r &     23.52 &     0.12 & GROND \\
        59635.26 &  7.99 &    i &     23.31 &     0.20 & GROND \\
        59635.26 &  7.99 &    J &   > 20.77 &          & GROND \\
        59635.26 &  7.99 &    H &   > 19.82 &          & GROND \\
        59635.26 &  7.99 &    K &   > 17.66 &          & GROND \\
        59639.28 &  12.00 &  gamma-ray &    &          & Swift (BAT)\\
        59639.28 &  12.00 &  xray & >8.1$\times10^{40}$ erg/s & & Swift (XRT) \\
        59639.28 &  12.00 &  uvm2 &   >21.78 &          & Swift (UVOT)\\
        59640.86 &  13.59 &   C  &    > $\sim$18-19 &   & Zadko \\
        59649.25 &  21.98 &    g &     23.12 &     0.09 & GROND \\
        59649.25 &  21.98 &    r &     23.87 &     0.17 & GROND \\
        59649.25 &  21.98 &    i &     23.87 &     0.41 & GROND \\
        59649.25 &  21.98 &    J &   > 20.52 &          & GROND \\
        59649.25 &  21.98 &    H &   > 19.55 &          & GROND \\
        59649.25 &  21.98 &    K &   > 17.63 &          & GROND \\
        59657.31 &  30.04 &    g &   > 22.93 &          & GROND \\
        59657.31 &  30.04 &    r &     23.49 &     0.37 & GROND \\
        59657.31 &  30.04 &    i &   > 23.13 &          & GROND \\
        59657.31 &  30.04 &    J &   > 20.43 &          & GROND \\
        59657.31 &  30.04 &    H &   > 19.52 &          & GROND \\
        59657.31 &  30.04 &    K &   > 17.06 &          & GROND \\
        59663.27 &  36.00 &    g &     23.15 &     0.14 & GROND \\
        59663.27 &  36.00 &    r &     23.52 &     0.11 & GROND \\
        59663.27 &  36.00 &    i &     23.69 &     0.19 & GROND \\
        59663.27 &  36.00 &    J &   > 20.80 &          & GROND \\
        59663.27 &  36.00 &    H &   > 19.66 &          & GROND \\
        59663.27 &  36.00 &    K &   > 17.33 &          & GROND \\
        59674.23 &  46.95 &    g &     22.92 &     0.08 & GROND \\
        59674.23 &  46.95 &    r &     23.52 &     0.12 & GROND \\
        59674.23 &  46.95 &    i &     23.43 &     0.19 & GROND \\
        59674.23 &  46.95 &    J &   > 20.69 &          & GROND \\
        59674.23 &  46.95 &    H &   > 19.73 &          & GROND \\
        59674.23 &  46.95 &    K &   > 17.53 &          & GROND \\
        59684.21 &  56.94 &    g &   > 22.83 &          & GROND \\
        59684.21 &  56.94 &    r &   > 23.45 &          & GROND \\
        59684.21 &  56.94 &    i &   > 23.08 &          & GROND \\
        59684.21 &  56.94 &    J &   > 20.55 &          & GROND \\
        59684.21 &  56.94 &    H &   > 19.43 &          & GROND \\
        59684.21 &  56.94 &    K &   > 17.65 &          & GROND \\
        \hline
    \end{tabular}
    \label{tab:AT2022kak_photometry}
\end{table}

\subsubsection{Radio}

We observed the position of AT2022kak in late-time with the Australia Telescope Compact Array \citep[ATCA;][]{Wilson:2011} on 2024-05-14, obtaining approximately 40 minutes on-source across the duration of a longer observation. Observations were carried out in the standard continuum mode with the 4cm receiver, with $2\times 2048$MHz bands centered on 5.5 and 9\,GHz. We calibrated and imaged the data using {\sc Miriad} \citep{1995ASPC...77..433S} and formed a single image in each band. We detect no source at the position of AT2022kak in either band, with $5\sigma$ upper limits of 110 and 190$\mu$Jy at 5.5 and 9\,GHz respectively.

\subsection{Spectroscopy}
\label{ssec:spectroscopy}

We performed spectroscopic follow-up on the location of AT2022kak after the first burst. Due to the rapid fading nature of the transient, spectroscopic follow-up was not acquired immediately after the discovery burst as it became too faint for ToO triggers on available follow-up facilities. We present here late-time follow-up from SOAR, AAT, and SALT. Fortunately, we gathered spectroscopic data while another burst was occurring as shown in Section \ref{AAT}.

\subsubsection{Southern Astrophysical Research Telescope (SOAR)}
\label{SOAR}

On 05 January 2025, the 4.1-meter Southern Astrophysical Research (SOAR) telescope was used to obtain the first spectroscopic data of the persistent source with the Goodman High throughput spectrograph \citep[GHTS;][]{Clemens:2004}. Eight exposures of 600s were taken with a grating of 400 lines/mm and a 1.0'' wide slit mask in the M1 spectroscopic setup (hereafter 400M1) with $2 \times 2$ binning using the GHTS Red Camera. The 400M1 spectra cover a wavelength range of 3800 -- 7040 \text{\AA}. 

The spectra were reduced using {\sc pypeit} \citep{pypeit:joss_arXiv, pypeit:zenodo}, using arcs taken immediately before and/or after target observation and calibration images from the same night. Flux calibration was performed using standard stars observed on the night of the observations with an identical 400M1 setup and $2 \times 2$ binning. We stacked the images using {\sc SWarp} \citep{Bertin:2002}, however due to the faintness of the persistent source, no apparent trace was detected in the stacked 2D spectra. 

\subsubsection{Anglo-Australian Telescope (AAT)}
\label{AAT}

We observed the persistent source with the Anglo-Australian Telescope \citep[AAT;][]{Lewis:2002} on February 25--27, 2025 using the Kilo-fibre Optical AAT Lenslet Array \citep[KOALA;][]{Ellis:2012} instrument on the  with resolution of R$\sim$1000. The blue arm used a $580V$ grating to cover $3635 - 5710$\text{\AA}, enabling detection of crucial emission and absorption features - including H$\beta$, H$\gamma$, H$\delta$, He I, He II, and Ca II lines - that characterise accretion processes and outburst phenomena in ultrafast dwarf novae such as AT2022kak. 
However, on the first night of observing, the source was discovered to be brighter than seen in the KNTraP template image, and thus determined to be in an outbursting state. As a result, the KOALA spectra are of the source in outburst through rise and fade.

On nights 1 and 3, the KOALA `wide-mode' configuration was used, which assigns each fibre a scale of 1.25$''$, and on night 2 the `narrow-mode' was used, which assigns each fibre 0.7$''$. The `wide-mode' configuration was chosen to maximise SNR and, as we noticed the source in outburst on night 2, we attempted to observe with higher resolution in night 2 while the source was bright and near peak.

The reliability of the baseline flux level throughout the observations may have been affected by several cloud interruptions throughout the observations, and additional possible extinction from a nearby bushfire on the third night. These environmental effects should be taken into consideration when viewing the flux levels of the spectra presented here. However, we are confident that the source shows outburst in the first night, as it had brightened relative to neighbouring stars in the field. 


A custom in-house data reduction pipeline, called {\sc PyKoala}, was used to process the spectroscopic KOALA data (note this code is not the AAO {\sc PyKOALA} software). This pipeline incorporates wrappers to run the standard 2dFDR reduction pipeline \citep{croom20042dfdr, sharp2010optimal} provided by the AAO. These wrappers automate the execution of standard basic reduction steps, including bias and dark subtractions, flat fielding, wavelength calibrations, and spectral extraction. Calibrations specific to the KOALA integral field unit (IFU) were developed to calibrate the flux by using a spectrophotometric standard star from the European Southern Observatory (ESO) spectrophotometric standard star list\footnote{\url{https://www.eso.org/sci/observing/tools/standards/spectra/stanlis.html}} and to perform an illumination calibration by using twilight flats. Issues with the pipeline's sky subtraction steps meant that this process had to be completed by hand. For each exposure, seven fibres forming a circular region were selected in an empty region of sky (same number and configuration as source fibres). A median was taken of these fibres to use as the sky background to subtract from the source fibres. In an effort to extract 1D spectra with maximised SNR when considering the seeing FWHM, the central fibre located at the AT2022kak location and surrounding six fibres (seven total fibres) were stacked together. 

\subsubsection{Southern African Large Telescope (SALT)}
\label{SALT}

Target of Opportunity observations were triggered during the second day of the 2025 outburst using the Robert Stobie Spectrograph (RSS) on the Southern African Large Telescope \citep[SALT;][]{Buckley:2006, O'Donoghue:2006} in a low-resolution mode (R~400-1200) with a total exposure time of 2400 s. Spectral extraction and wavelength calibration were performed using standard packages in IRAF. Relative flux calibration was carried out using sensitivity functions derived from observations of the spectrophotometric standard star HD 289002. The standard star was observed with the same grating settings as the target, using an exposure time of 30 seconds.

\section{Analysis}
\label{sec:analysis}

\subsection{Light Curve Characterisation}
\label{LCchar}

\begin{figure}
    \centering
    \includegraphics[width=1.0\linewidth]{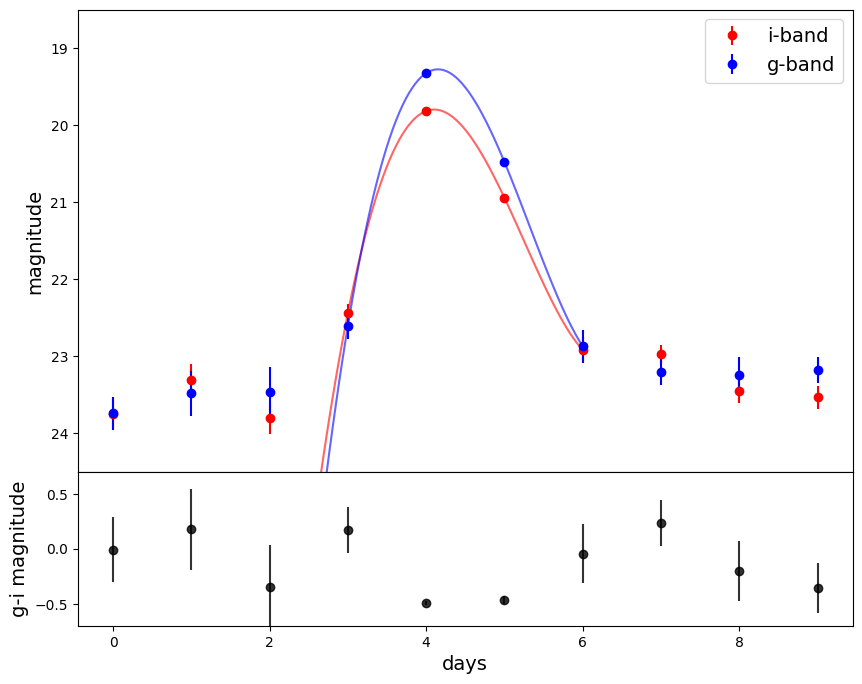}
    \caption{Top: Light curve of the February 2022 outburst and best-fit polynomials of fourth order. The polynomial fit is only applied to outburst nights 3 -- 6. Bottom: Light curve (\emph{g-i}) colour evolution.}
    \label{fig:polynomialfit_colourev}
\end{figure}

\begin{figure}
    \centering
    \includegraphics[width=1\linewidth]{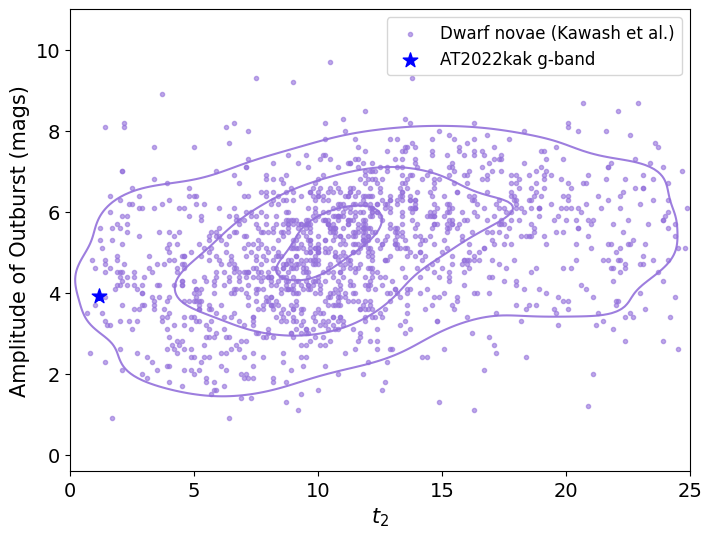}
    \caption{Distribution of dwarf novae plotted as a function of their amplitude of outburst in magnitudes versus $t_2$ in days. The $t_2$ value of AT2022kak in the \emph{g}-band is indicated with a star marker. Contours indicate 10\%, 50\%, and 90\% density levels. Data is taken from \citet{Kawash:2021} and magnitudes are measured in ASAS-SN \emph{g}-band.}
    \label{fig:dwarfnovae_dist}
\end{figure}

Polynomial fitting was applied to the photometric light curve of the 2022 outburst, shown in Figure \ref{fig:polynomialfit_colourev}. A fourth order polynomial was used to characterise the outburst data points (nights 3 -- 6 inclusive). From this fit, we find a peak magnitude of $m_g$ = 19.27 and $m_i$ = 19.80 and rise and fade rates of $\Delta m_{g(i)}=$ 3.4 (3.35) mag day$^{-1}$ and $\Delta m_{g(i)}=$ 2.12 (1.95) mag day$^{-1}$ respectively. These value assume a quiescent magnitude of $m_g$ = 23.19 and $m_i$ = 23.49, as calculated from the quiescent DECam and GROND median magnitudes. We also calculate $t_2$, defined as the time in days for the burst to fade two magnitudes from peak. Using the fit, we find $t_2$ = 1.16\,d in \emph{g}-band, and $t_2$ = 1.28\,d in \emph{i}-band. A comparison of these $t_2$ values to the population of DNe is shown in Figure \ref{fig:dwarfnovae_dist}. Table \ref{tab:lightcurve_properties} lists the polynomial fit results. 

The \emph{g--i} colour evolution of AT2022kak is shown on the bottom panel of Figure \ref{fig:polynomialfit_colourev}. The event is blue during outburst and becomes redder near the end of its fade. Its \emph{(g-i)} = $-$0.49 colour at peak is typical of DN outbursts \citep{Cannizzo:1987}.

To check for recurrent historical outbursts, several online catalogues were queried. We generated a light curve on the Asteroid Terrestrial-impact Last Alert System \citep[ATLAS;][]{Tonry:2018} forced photometry server \citep{Shingles:2021} from February 29, 2016 to present day, and is shown in Appendix \ref{appsec:atlas_xmatch}. The light curve appears to show four bursts over this time-frame, however we inspected the ATLAS images and found all the detections to be spurious. Additionally, we searched The Transiting Exoplanet Survey Satellite \citep[TESS;][]{Ricker:2015}, SkyMapper Telescope and the Southern Sky Survey \citep{Keller:2007}, and All-Sky Automated Survey for SuperNovae \citep[ASAS-SN;][]{Shappee:2014} catalogs. No cross-matched detections were found in these catalogs. However, we note that any past outburst with similar or fainter brightness compared to the 15 February 2022 outburst would be missed by these surveys due to their detection limits. 

\begin{table}
    \centering
    \caption{Polynomial fit light curve parameters.}
    \begin{tabular}{l|c|c}
        \hline
        Parameter & \emph{g}-band & \emph{i}-band \\
        \hline
        Peak magnitude (mag)        & 19.27 & 19.80 \\
        Rise rate (mag day$^{-1}$)  & 3.40  & 3.35  \\ 
        Fade rate (mag day$^{-1}$)  & 2.12  & 1.95  \\ 
        Amplitude of outburst (mag) & 3.92  & 3.70  \\
        $t_2$ (days)                & 1.16  & 1.28  \\
        \hline
    \end{tabular}
    \label{tab:lightcurve_properties}
\end{table}

Some DNe are seen to also have `superoutbursts', which are typically brighter by $\sim$0.8 mag in \emph{V}-band, and last longer than normal outbursts, with typical outburst durations of $\sim$12--20 d \citep{Patterson:2011, Otulakowska-Hypka:2016}. As survey searches above have a higher probability of detecting a superoutburst, the likelihood of AT2022kak being classified as a SU UMa DN type, which have known regular superoutbursts, is low. 






\begin{table}
    \centering
    \caption{Physical parameters of AT2022kak estimated using tables from \citet{Knigge:2011} and an orbital period of 1.4$\pm0.2$ hr.}
    \begin{tabular}{l|l}
        \hline
        Parameter & Value \\
        \hline
        Orbital period                   & 1.4$\pm0.2$hr\\
        Mass of primary star             & 0.75 $M_\odot$ \\
        Mass of secondary star           & 0.089 $M_\odot$\\
        Radius of secondary star         & 0.13 $R_\odot$\\
        Distance modulus                 & 13.9$\pm0.2$ mag\\
        Distance from Sun                & 6.2$\pm0.5$ kpc \\
        Distance from Galactic centre    & 6.6$\pm0.1$ kpc \\
        Scale height from Galactic plane & 1.9$\pm0.2$ kpc\\
        \hline          
    \end{tabular}
    \label{tab:AT2022kak_physicalparameters}
\end{table}

\begin{figure*}
    \centering
    \begin{subfigure}
        \centering
        \includegraphics[width=0.9\textwidth]{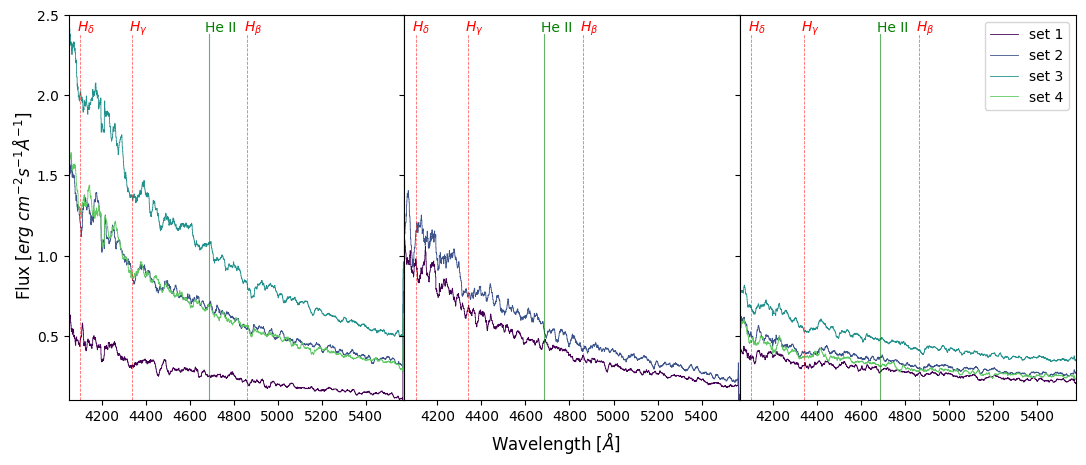}
    \end{subfigure}

    \begin{subfigure}
        \centering
        \includegraphics[width=0.9\textwidth]{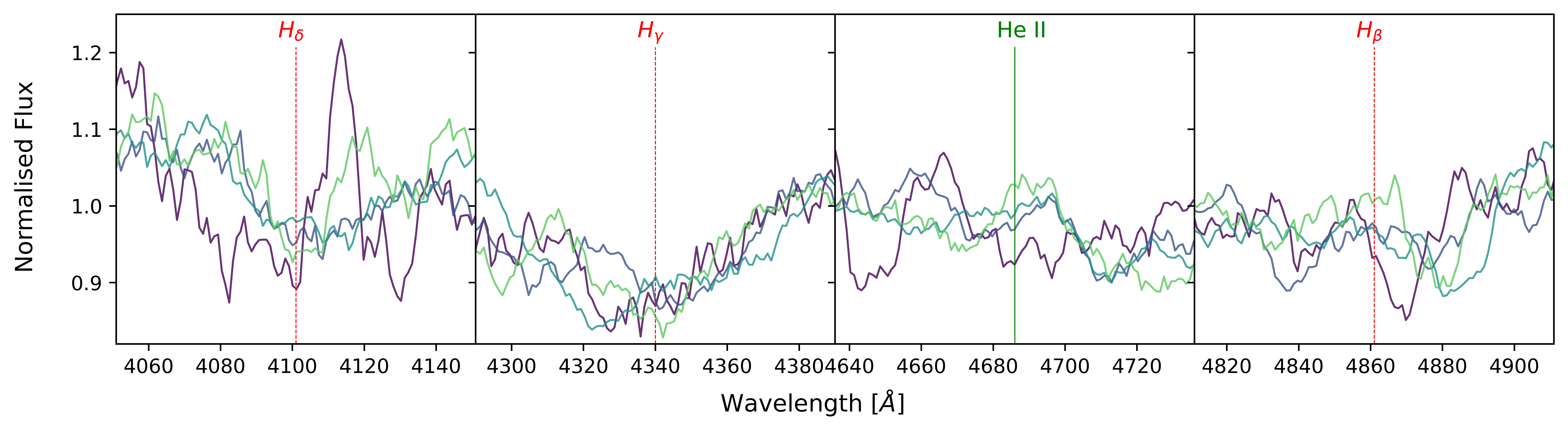}
    \end{subfigure}

    \centering
    \caption{(\emph{Top}) Panels from left to right shows KOALA spectra in multiple stacked sets to show nightly evolution from observing nights one, two, and three (MJD 60731.56, 60732.55, 60733.51). Night one and three have three 20 minute exposures per stacked set, except set 1 night one, which has two. Night two has a stacks of two and three exposures for sets 1 and 2 respectively. (\emph{Bottom}) Stacked and normalised spectra from night one, showing four zoomed in features expected of a DN; Balmer series ($H_{\beta}, H_{\gamma}, H_{\delta}$) and He II. Vertical lines expected to be absorption DN features marked with dashed lines, emission features marked with solid lines.}
    \label{fig:koala_spectra_and_features}
    
\end{figure*}

\subsection{Spectral Analysis}

\begin{figure}
    \centering
    \includegraphics[width=1\linewidth]{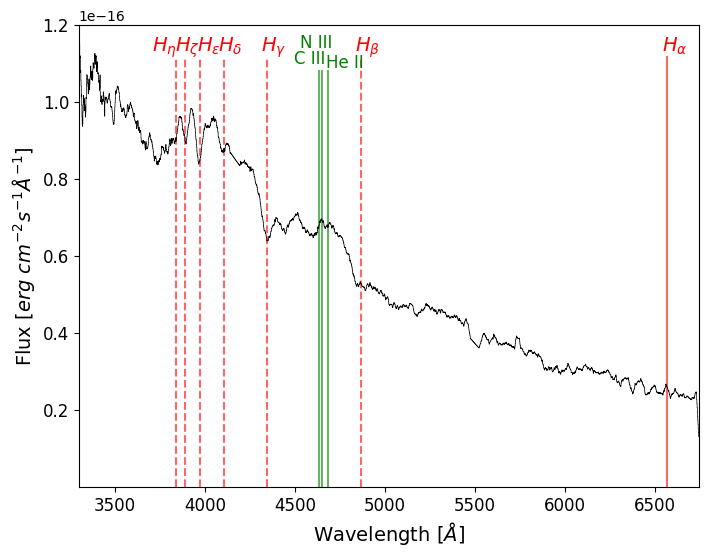}
    \caption{Spectrum of AT2022kak during the 26 February 2025 outburst with SALT. DN absorption and emission features labelled same as in \ref{fig:koala_spectra_and_features}.}
    \label{fig:salt_spectra}
\end{figure}

The low-resolution 2025 outburst spectra collected by SALT and KOALA show Balmer absorption lines, which are expected for DNe, however the SNR is too low to confirm weak H$\alpha$ emission (Figures \ref{fig:koala_spectra_and_features} and \ref{fig:salt_spectra}). The He II $\lambda$4686 feature is common for DNe, however it is not seen strongly in the KOALA spectra.  Generally, these features are characteristic of DNe outburst spectra, consistent with several documented outburst spectra in the literature (e.g. AR And-20160105, RZ Lmi-20131112, V344 Lyr-20160507 \citep{Han:2020} and several DNe presented in \citet{Morales-Rueda:2002}). A potential very broad emission feature is seen in the SALT spectrum, possibly a blend with the C III/ N III $\lambda$4634-4651 features, which may be indicative of some structure in the accretion disk. However, the broadening is likely too wide to be physically realistic for a DNe.

\begin{figure}
    \centering
    \includegraphics[width=1\linewidth]{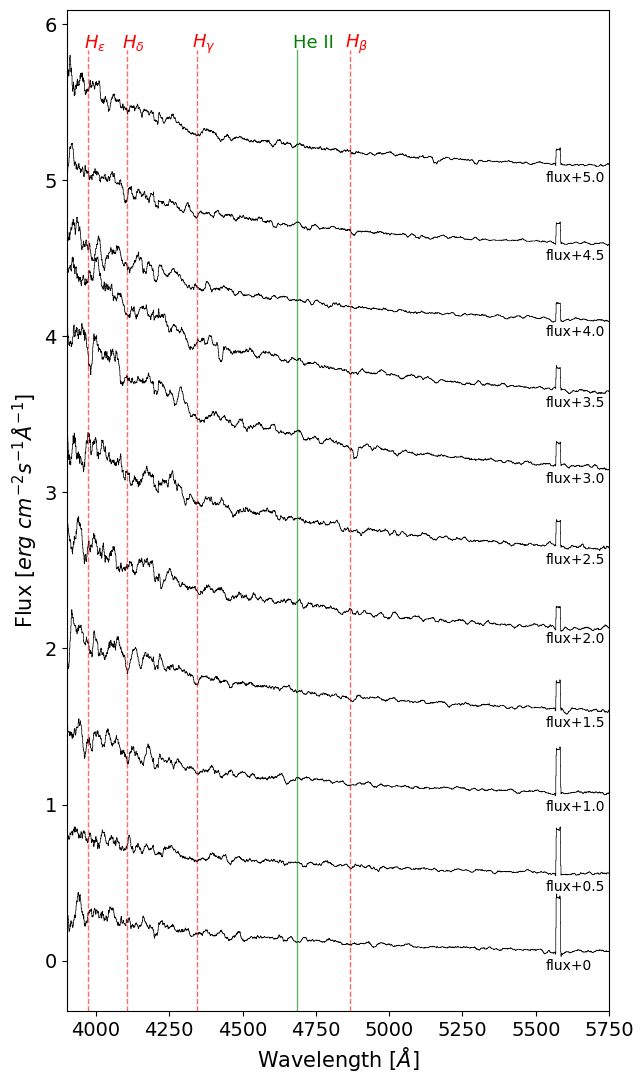}
    \caption{Spectral evolution of AT2022kak on 25 February 2025. The individual 20\,m exposures are shown in chronological order from bottom to top with flux offsets to help with clarity. The outburst appears to begin by the fourth exposure. DN absorption and emission features are labelled similarly to \ref{fig:koala_spectra_and_features}. Note that the feature at $\sim\lambda$5600 is an artefact from the sky background.}
    \label{fig:KOALA_night1_spectra}
\end{figure}

As described in Section \ref{ssec:spectroscopy}, the relative flux levels may be affected by occasional clouds and experience reddening from possible smoke extinction from a fire nearby the observatory. However, relative to other sources in the field, it is clear that the event is caught in outburst on night 1 (Figure \ref{fig:koala_spectra_and_features} upper panels and Figure \ref{fig:KOALA_night1_spectra}). 

The unstacked night 1 exposures in Figure \ref{fig:KOALA_night1_spectra} show $\sim$20\,m spectral time resolution of the outburst rising to peak. The night 1 DN features (Balmer series, He II) are shown in detail in the bottom panels of \ref{fig:koala_spectra_and_features}. As the spectra have relatively low SNR, no conclusive comments can be made regarding feature evolution during outburst.



\section{Results}

\subsection{Dwarf Nova Properties}

We search for a potential orbital period by using a statistical study done by \cite{Otulakowska-Hypka:2016}, which finds a relation correlating orbital period and outburst duration. The best fit of this correlation is shown in Equation (19) of that study:
\begin{equation}
    log D_{NO} = 0.745(\pm0.060) \cdot log P_{orb} + 0.398(\pm0.033)
\end{equation}

where $D_{NO}$ is the outburst duration in days, from when the outburst begins to when it returns back to quiescence. We estimate the duration of the AT2022kak 2022 outburst to be 3.2$\pm{0.3}$ days using a Monte Carlo method, thus giving an orbital period of 1.4$\pm0.2$ hr.

Using this orbital period, and model CV parameter tables from \cite{Knigge:2011}, further physical parameters are estimated and summarised in Table \ref{tab:AT2022kak_physicalparameters}. 

In the context of DNe from \citet{Kawash:2021}, comparing $t_2$ and amplitude of outburst in Figure \ref{fig:dwarfnovae_dist}, the distribution shows that AT2022kak is one of the fastest fading DNe seen to date. However as noted, we do not know the true peak of this outburst due to the cadence of observations, thus the $t_2$ value may be faster than reported here. In this subset of the full catalog from \citet{Kawash:2021}, where only DNe with $t_2 < 25$ days are shown, AT2022kak is in the top $\sim$0.3\% of fastest DNe.  

DNe fading on these ultrafast timescales can occur when a magnetic WD is part of the binary system; these systems are known as intermediate polars \citep[see][for a review]{Patterson:1994}. An example of an intermediate polar is EX Hya \citep{Reinsch:1990}, which displays similar light curve, with a rise to peak of $\lesssim$ 12h, fade to quiescence of $\sim$2 days, and outburst amplitude $\sim$3 mags.

The morphology of the light curve shows a relatively symmetric shape with the polynomial fit. This could be indicative of an `inside-out' outburst, where in accordance with the disk-instability model, which currently is the best model to describe DN outbursts \citep{Osaki:1996, Lasota:2001}, the outburst is initially triggered on the inner section of the accretion disk, before propagating both inwards and outwards \citep{Mineshige:1985}. However, due to the cadence of observations, we cannot tell if the true peak of the outburst occurs before the fourth night, which would give the burst a more asymmetric shape. 

\subsection{Distance and Galactic Location}

Given that KNTraP observed fields outside of the Galactic plane, we study the potential location of this DN. We estimate the peak absolute magnitude of AT2022kak to estimate its distance. We use Equation (3) of \citet{Patterson:2011}:
\begin{equation}
    M_{max} = 5.70 (\pm0.18) - 0.287 (\pm 0.018) P_{orb}
\end{equation}

where $P_{orb}$ is the orbital period, developed from a known range of DN peak absolute magnitudes for short period CVs. Using our orbital period estimate (\S\ref{LCchar}) and this equation, we estimate a $M_V$ = 5.3$\pm0.2$ peak absolute magnitude for AT2022kak. 

Using the distance modulus, AT2022kak is estimated to be located at a distance of 6.2$\pm0.5$ kpc from Earth. The galactic coordinates place AT2022kak 6.6$\pm0.1$ kpc from the Galactic centre, and at a height of 1.9$\pm0.2$ kpc above the Galactic plane. 

A distance lower limit is estimated by assuming a faint secondary star. V388 Cas, a prototypical M5 type star is used \citep{Alonso_Floriano:2015}, as M dwarfs are common secondary types. The magnitude of this prototypical star ($M_G=11.9$) does not include contributions of the WD or accretion disk in the system, therefore the absolute magnitude of the system must be brighter. By only using a faint type of secondary star (as other systems may have brighter secondaries), this estimate places a lower limit on the distance of the system. 

We calculate a lower limit of the distance to AT2022kak to be $>$1.8 kpc from Earth, $>$7.3 kpc from the Galactic centre, and $>$0.6 kpc above the Galactic plane. The estimated distances of AT2022kak place it outside the thin disk, and potentially outside the thick disk, where most DNe are found. 

This is interesting as CVs with Population II secondaries are estimated to be found at heights $\gtrsim$ 2 kpc above the Galactic plane \citet{Stehle:1997} found. Additionally, Population II CVs are expected to have orbital periods below the $\sim$2-3 hr gap in the observed DN period distribution as determined by population synthesis models \citep{Verbunt:1981, Rappaport:1982, Rappaport:1983, Paczynski:1983, Spruit:1983, King:1988, Shao:2012}. As AT2022kak is estimated to have a height of $1.9\pm0.2$ kpc, and orbital period of $1.4\pm0.2$ hr, it is a thick disk candidate, and potential Population II DN.

\section{Conclusions}
\label{sec:conclusion}

We discovered a fast fading optical transient, AT2022kak, with KNTraP, observed to rise to peak by $\Delta m_g >$ 3.87 and $\Delta m_i >$ 3.68 from the calculated median quiescent magnitude, and fade by the same amount in two nights. We performed a multi-wavelength follow-up campaign over two months where no other variability was observed. In February 2025 we obtained spectroscopic data coincidentally at the beginning of a recurrent burst. High time resolution (20 min) spectra was captured over three nights through the rise, peak, and fade of the outburst. 

The light curve shape and properties of the initial 2022 outburst, as well as the spectral features and evolution seen in the 2025 outburst, all consistent with dwarf novae. Main results from this work include:

\begin{itemize}
    \item AT2022kak is found to be one of the fastest (top $<$0.3\%) fading dwarf novae, with a $t_2$ of 1.16 days in the \emph{g}-band, and 1.28 days in the \emph{i}-band, as calculated from the polynomial fits
    \item AT2022kak is distant and estimated to be located $1.9\pm0.2$2 kpc above the Galactic plane, making it a potential Population II DN.
\end{itemize}

AT2022kak adds to the number of candidate Population II dwarf novae, which will help uncover the underlying accretion disk mechanics. Further observations of this dwarf nova will be beneficial to understand the nature of the system components and accretion disk, and place better constraints on its physical parameters, such as orbital period and recurrence time.

Deep, day-cadenced surveys like KNTraP are uncovering a new sample space in the transient domain. AT2022kak is an example of a faint, rapidly evolving transient which could not have been identified with other existing transient surveys. Future transient facilities coming online in the near future such as the Vera C. Rubin Observatory \citep{Ivezić:2019} and Nancy Grace Roman Space Telescope \citep{Spergel:2015} will reach similar and/or deeper depths over larger areas compared to KNTraP, and also will uncover these intrinsically fast and faint transients. However, unlike Rubin and Roman, KNTraP's fast day cadence observations allows for better sampling and thus characterisation of transients on AT2022kak timescales. Therefore, KNTraP and KNTraP-like surveys will be important to help with the identification of the new transients in this faint and fast parameter space that will be explored with Roman, and the upcoming Rubin Legacy Survey of Space and Time.



\section*{Acknowledgements}

This research was funded in part by the Australian Research Council Centre of Excellence for Gravitational Wave Discovery (OzGrav), CE170100004 and CE230100016. 

JC acknowledges funding from the Australian Research Council Discovery Project DP200102102. 

AM is supported by DE230100055. 

Part of the funding for GROND (both hardware as well as personnel) was generously granted from the Leibniz-Prize to Prof. G. Hasinger (DFG grant HA 1850/28-1).

This work has made use of data from the Asteroid Terrestrial-impact Last Alert System (ATLAS) project. The Asteroid Terrestrial-impact Last Alert System (ATLAS) project is primarily funded to search for near earth asteroids through NASA grants NN12AR55G, 80NSSC18K0284, and 80NSSC18K1575; byproducts of the NEO search include images and catalogs from the survey area. This work was partially funded by Kepler/K2 grant J1944/80NSSC19K0112 and HST GO-15889, and STFC grants ST/T000198/1 and ST/S006109/1. The ATLAS science products have been made possible through the contributions of the University of Hawaii Institute for Astronomy, the Queen’s University Belfast, the Space Telescope Science Institute, the South African Astronomical Observatory, and The Millennium Institute of Astrophysics (MAS), Chile.

\section*{Data Availability}

 Calibrated DECam images from this survey are publicly available on the NOIRLab Astro Data Archive under program number 2022A-679480. The web portal can be viewed at \url{https://astroarchive.noirlab.edu}.



\bibliographystyle{mnras}
\bibliography{references} 




\appendix

\section{Spectral Energy Distribution}
\label{appsec:at2022kak_sed}

\begin{figure}
    \centering
    \includegraphics[width=1\linewidth]{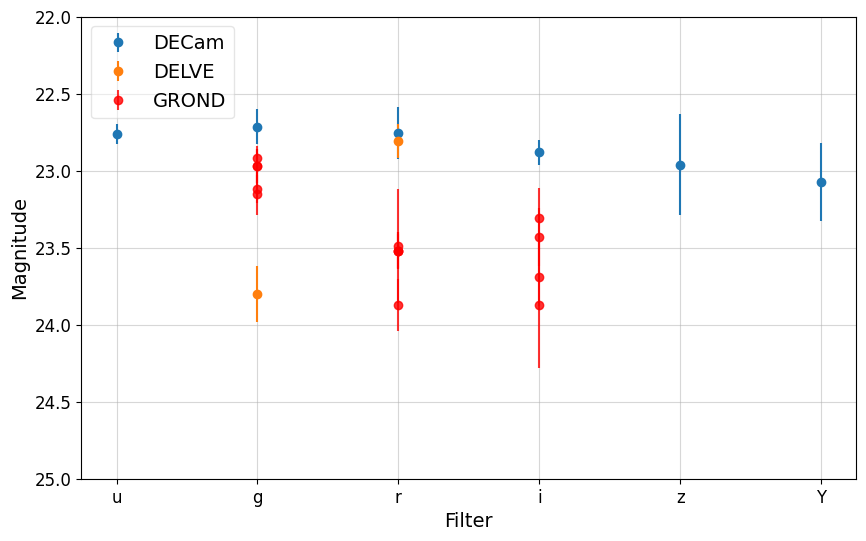}
    \caption{A course SED with combined DECam data from KNTraP (\emph{g, i}), and archival data from NOIRLab (\emph{u, r, z, Y}). Included also are the \emph{g} and \emph{r} data points from the DELVE catalog, and GROND photometry in \emph{g, r} and \emph{i} bands.}
    \label{fig:rough_sed}
\end{figure}

In figure \ref{fig:rough_sed} we present a spectral energy distribution of archival and KNTraP DECam photometry, DELVE catalog photometry, and GROND photometry. The total exposure times for stacked images in each filter as follows; \emph{u}: 2400s, \emph{g}: 2100s, \emph{r}: 520s, \emph{i}: 2550s, \emph{z}:460s, and \emph{Y}: 120s.

\section{ATLAS Crossmatch}
\label{appsec:atlas_xmatch}

\begin{figure*}
    \centering
    \includegraphics[width=1\textwidth]{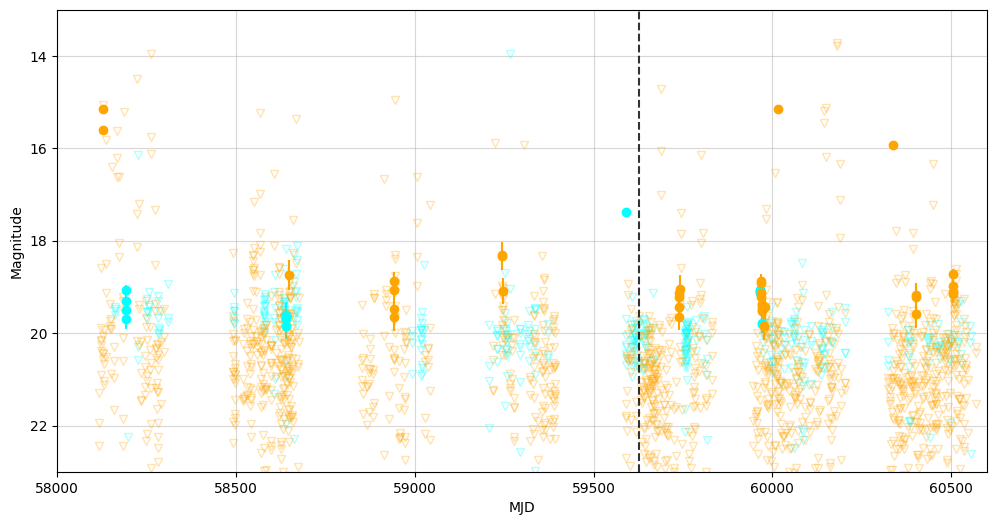}
    \caption{ATLAS forced photometry light curve at the location of AT2022kak, in the orange and cyan filters. Detections are shown as full markers, and upper limits and detections with a SNR $<$ 3 are shown as hollow inverted triangles. The vertical dotted line indicates the time of the AT2022kak burst as detected with KNTraP.}
    \label{fig:atlas_lc}
\end{figure*}

In figure \ref{fig:atlas_lc} we present the ATLAS forced photometry light curve at the location of AT2022kak. From visual inspection of the image cutouts, we determine that the four apparent bursts in this light curve to be spurious detections.

\bsp	
\label{lastpage}
\end{document}